\documentclass[useAMS]{mn2e}

\usepackage{graphicx}

\usepackage{rotating}
\newcommand\kms{{\rm\,km\,s^{-1}}}

\newcommand\msun{\rm\,M_\odot}
\newcommand\rsun{\rm\,R_\odot}

\def\apgt{\ {\raise-.5ex\hbox{$\buildrel>\over\sim$}}\ }
\def\aplt{\ {\raise-.5ex\hbox{$\buildrel<\over\sim$}}\ }

\title[Very massive runaway stars from three-body encounters]{Very massive runaway stars from three-body encounters}
\author[V.V.Gvaramadze and A.Gualandris]
       {Vasilii V. Gvaramadze$^{1,2}$\thanks{E-mail: vgvaram@mx.iki.rssi.ru} and
       Alessia Gualandris$^{3}$\thanks{E-mail: alessiag@mpa-garching.mpg.de}\\
       $^{1}$Sternberg Astronomical Institute, Moscow State University,
       Universitetskij Pr. 13, Moscow 119992, Russia\\
       $^{2}$Isaac Newton Institute of Chile, Moscow Branch,
       Universitetskij Pr. 13, Moscow 119992, Russia\\
       $^{3}$Max-Planck Institut f\"{u}r Astrophysik,
       Karl-Schwarzschild-Str. 1, D-85741 Garching, Germany\\
       }
\begin{document}

\date{Accepted 2010 July 28. Received 2010 July 27; in original form 2010 June 25}

%\pagerange{\pageref{firstpage}--\pageref{lastpage}} \pubeyar{2009}

\maketitle

\begin{abstract}
Very massive stars preferentially reside in the cores of their
parent clusters and form binary or multiple systems. We study the
role of tight very massive binaries in the origin of the field
population of very massive stars. We performed numerical
simulations of dynamical encounters between single (massive) stars
and a very massive binary with parameters similar to those of the
most massive known Galactic binaries, WR\,20a and NGC\,3603-A1. We
found that these three-body encounters could be responsible for
the origin of high peculiar velocities ($\geq 70 \, {\rm km} \,
{\rm s}^{-1}$) observed for some very massive ($\geq 60-70 \,
M_{\odot}$) runaway stars in the Milky Way and the Large
Magellanic Cloud (e.g., $\lambda$\,Cep, BD+43$\degr$\,3654,
Sk\,$-$67$\degr$22, BI\,237, 30\,Dor\,016), which can hardly be
explained within the framework of the binary-supernova scenario.
The production of high-velocity massive stars via three-body
encounters is accompanied by the recoil of the binary in the
opposite direction to the ejected star. We show that the relative
position of the very massive binary R145 and the runaway early
B-type star Sk$-$69$\degr$\,206 on the sky is consistent with the
possibility that both objects were ejected from the central
cluster, R136, of the star-forming region 30\,Doradus via the same
dynamical event -- a three-body encounter.
\end{abstract}


%---------------------------------------------------------------------
\section{Introduction}
%---------------------------------------------------------------------
%

There is growing observational evidence that most (and possibly
all) massive stars are formed in clusters (Lada \& Lada 2003; cf.
de Wit et al. 2005; Schilbach \& R\"{o}ser 2008; Gvaramadze \&
Bomans 2008b) and that all (or most) O-type stars (either in
clusters or in the field) are (or were) members of binary or
multiple systems (Mason et al. 1998; Preibisch, Weigelt \&
Zinnecker 2001; Garc\'{i}a \& Mermilliod 2001; Kobulnicky \& Fryer
2007; Clark et al. 2008). Observations also show that the binary
frequency increases with stellar mass (Larson 2001; Clark et al.
2008) and that the most massive binaries are usually the most
short-period (tight) ones (e.g. Mermilliod \& Garc\'{i}a 2001).
Moreover, a high proportion of massive binaries have mass ratios
close to unity (Clarke \& Pringle 1992; Pinsonneault \& Stanek
2006; Kobulnicky \& Fryer 2007) and this proportion is larger in
close binaries (Mason et al. 1998). The best examples of very
massive and tight binaries with companions of comparable mass are
the WN6ha binary systems WR\,20a ($83\msun + 82\msun$; Bonanos et
al. 2004; Rauw et al. 2005) and NGC\,3603-A1 ($116\msun +
89\msun$; Schnurr et al. 2008a).

The WR\,20a binary is also remarkable by its significant
displacement ($\sim 1$ pc) from the centre of the parent cluster
Westerlund\,2. This displacement strongly suggests that the binary
experienced a dynamical encounter with another massive star
(either single or binary) and consequently recoiled (cf. Rauw et
al. 2005; see also Section\,\ref{sec:dis}). The purpose of this
paper is to study numerically the role of three-body dynamical
encounters between tightly bound (hard; Aarseth \& Hills 1972;
Hills 1975; Heggie 1975) binaries and a third star in the
production of very massive runaway stars (e.g., $\lambda$\,Cep,
BD+43$\degr$\,3654, Sk\,$-$67$\degr$22, BI\,237, 30\,Dor\,016; see
Section\,\ref{sec:VMR} for a summary of these stars) whose high
peculiar velocities can hardly be explained within the framework
of the binary-supernova scenario (Section\,\ref{sec:binarySN}). In
Section\,\ref{sec:hills} we discuss the dynamical ejection
scenario. In Section\,\ref{sec:num} we present the results of our
numerical experiments. The obtained results are discussed in
Section\,\ref{sec:dis}.

\section{Very massive runaway stars}
\label{sec:VMR}

The stars with peculiar velocities exceeding $30 \, \kms$ are
called runaway stars (Blaauw 1961). The origin of these stars can
be attributed to two basic processes: (i) disruption of a
short-period binary system following the (asymmetric) supernova
explosion of one of the binary components (Blaauw 1961; Stone
1991) and (ii) dynamical three- or four-body encounters in dense
stellar systems (Poveda, Ruiz \& Allen 1967; Gies \& Bolton 1986).
Observations show that the percentage of runaway stars is highest
($\sim 25$ per cent) among the O stars and steeply decreases to
several per cent for the B stars and to even smaller values for
the less massive stars (e.g. Gies 1987; Blaauw 1993; Zinnecker \&
Yorke 2007). This tendency is consistent with the fact that the
massive stars prefer to reside in the cores of the parent clusters
(either due to dynamical or primordial mass segregation) where the
dynamical encounters between the cluster members are most frequent
and energetic, and suggests that the production of runaway stars
is dominated by the second process. On the other hand, in both
processes the less massive stars could be accelerated to larger
velocities, which is consistent with the observed anticorrelation
between the mass and the velocity of runaway stars (Gies \& Bolton
1986). The record-holder among the Galactic massive runaway stars
is the early B-type ($\simeq 11\, \msun$) star HD\,271791, whose
peculiar velocity ($\simeq 530-920 \, \kms$; Heber et al. 2008) is
about an order of magnitude larger than that of BD+43$\degr$\,3654
-- the fastest known Galactic early O-type star (the linear
momenta of both stars however are comparable).

BD+43$\degr$\,3654 is an O4If (Comer\'{o}n \& Pasquali 2007)
runaway star ejected from the Cyg\,OB2 association about 1.8 Myr
ago with a peculiar transverse velocity of $\simeq 40 \, \kms$
(Comer\'{o}n \& Pasquali 2007; cf. Gvaramadze \& Bomans 2008a).
Like many other runaway stars, BD+43$\degr$\,3654 generates a bow
shock visible in the infrared; see Van Buren \& McCray (1988) and
Comer\'{o}n \& Pasquali (2007) for the {\it Infrared Astronomical
Satellite (IRAS)} and the {\it Midcourse Space Experiment (MSX)}
satellite images of the bow shock, respectively. The heliocentric
radial velocity of BD+43$\degr$\,3654 of $-66 \pm 9 \, \kms$
(Kobulnicky, Gilbert \& Kiminki 2010) is much larger than the mean
systemic velocity of Cyg\,OB2 of $-10.3\pm 0.3 \, \kms$ (Kiminki
et al. 2007), which also supports the runaway nature of the star.
With an initial mass of $\simeq 70\pm 15 \, \msun$ (Comer\'{o}n \&
Pasquali 2007) and a total peculiar velocity of $\simeq 70 \,
\kms$, BD+43$\degr$\,3654 is the most massive known runaway star
in the Galaxy.

\begin{figure}
\includegraphics[width=8.5cm]{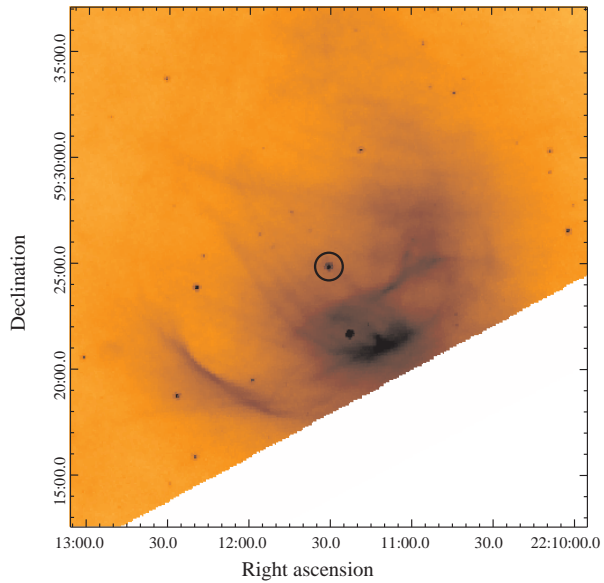}
\centering \caption{{\it Spitzer} $24\,\mu$m image of the bow
shock associated with the O6\,I(n)f runaway star $\lambda$ Cep.
The position of the star is indicated by a circle.}
\label{fig:Cep}
\end{figure}

Another example of very massive Galactic runaway stars is the
O6I(n)f (Walborn 1973) star $\lambda$\,Cep (HD\,210839). Like
BD+43$\degr$\,3654, $\lambda$\,Cep produces a bow shock,
originally discovered with {\it IRAS} by van Buren \& McCray
(1988). In Fig.\,\ref{fig:Cep} we present for the first time the
{\it Spitzer Space Telescope} 24\,$\mu$m image showing the fine
structure of the bow shock. (The image was retrieved from the {\it
Spitzer} archive using the Leopard software.) Using the parallax
and the proper motion from the new reduction of the {\it
Hipparcos} data (van Leeuwen 2007), one finds the peculiar
velocity of $\lambda$\,Cep in  Galactic coordinates: $v_l \simeq
-32 \, \kms$, $v_b \simeq -7 \, \kms$ [we used here the Galactic
constants $R_0 = 8.4$ kpc and $\Theta _0 =254 \, {\rm km} \, {\rm
s}^{-1}$ (Reid et al. 2009) and the solar peculiar motion
$(U_{\odot} , V_{\odot} , W_{\odot} )=(10.0, 11.0, 7.2) \, {\rm
km} \, {\rm s}^{-1}$ (McMillan \& Binney 2010)]. The orientation
of the bow shock and the direction of the transverse peculiar
velocity are consistent with the possibility that $\lambda$\,Cep
was ejected about 2.5 Myr ago from the Cep\,OB3 association,
located at $\simeq 6\fdg6$ to the east from the star (cf.
Hoogerwerf, de Bruijne \& de Zeeuw 2001). The current mass of
$\lambda$\,Cep is $\simeq 45-60 \, \msun$ (Martins, Schaerer \&
Hillier 2005; Repolust, Puls \& Herrero 2004), which implies that
the star was ejected very soon after its birth in the association.
Adopting the heliocentric radial velocity of $\lambda$\,Cep of
$\simeq -76 \, \kms$ (Conti, Leep \& Lorre 1977) and the systemic
velocity of Cep\,OB3 of $\simeq -23 \, \kms$ (Mel'nik \& Dambis
2009), one finds a total peculiar velocity for the star of $\simeq
60 \, \kms$.
\begin{table}
  \caption{Very massive runaway stars. For the first two (Galactic) stars we give their
  full (three-dimensional) peculiar velocities, while for the remaining
  %five
  four stars (located in the LMC) we give the peculiar radial velocities only.}
  \label{tab:VMR}
  \renewcommand{\footnoterule}{}
  \begin{center}
  \begin{minipage}{\textwidth}
\begin{tabular}{lcccccccccc}
\hline
Star & Sp. type & $v$ ($\kms$) & Association  \\
\hline
BD+43$\degr$\,3654 & O4If & $\simeq 70^{(1)}$ & Cyg\,OB2 \\
$\lambda$\,Cep & O6I(n)f & $\simeq 60^{(2)}$ & Cep\,OB3 \\
N11-026 & O2.5\,III(f*) & $\simeq 35^{(3)}$ & LH\,10 ? \\
Sk\,$-$67$\degr$22 & O2\,If* & $\simeq 150^{(4)}$ & ?\\
BI\,237 & O2\,V((f*)) & $\simeq 120^{(4)}$ & LH\,82 \\
30\,Dor\,016 & O2III-If* & $\simeq 85^{(5)}$ & 30\,Doradus \\
\hline
\end{tabular}
 \end{minipage}
    \end{center}
(1) Comer\'{o}n \& Pasquali 2007; Kobulnicky et al. 2010; (2) van
Leeuwen 2007; Conti et al. 1977; (3) Evans et al. 2006; (4) Massey
et al. 2005; (5) Evans et al. 2010.
%\end{minipage}
\end{table}

Finally, we note several very massive (O2-type) stars in the Large
Magellanic Cloud (LMC) whose high radial velocities ($\sim 40-150
\, \kms$ greater than the LMC systemic velocity) were interpreted
as an indication that these stars are runaways (Massey et al.
2005; Evans et al. 2006, 2010; cf. Nota et al. 1994; Danforth \&
Chu 2001; Schnurr et al. 2008b). A strong support for this
interpretation comes from the discovery of bow shocks associated
with one of these candidate runaway stars, BI\,237, and several
other OB stars in the field of the LMC (Gvaramadze, Kroupa \&
Pflamm-Altenburg 2010; see also Section\,\ref{sec:dis}). The most
striking runaway in the LMC is the O2III-If* star 30\,Dor\,016
(Evans et al. 2010), which is located in the periphery of the
star-forming complex 30\,Doradus, at $\simeq 8'$ ($\simeq 120$ pc
in projection) from R136 (the central cluster of 30\,Doradus) or
$\simeq 5'$ ($\simeq 70$ pc) from NGC\,2060 (another cluster in
30\,Doradus). The very large mass of 30\,Dor\,016 ($\sim 90 \,
\msun$; Evans et al. 2010) makes this star the most massive known
runaway.

The summary of the very massive runaway stars in the Galaxy and
the LMC is given in Table\,\ref{tab:VMR}. The last column
gives the birthplaces of the stars.

\section{Binary-supernova scenario}
\label{sec:binarySN}

We now consider whether or not the origin of very massive runaway
stars can be explained within the framework of the
binary-supernova scenario. According to this scenario, a runaway
star attains its peculiar velocity in the process of
disintegration of a binary system following the supernova
explosion of the primary (initially more massive) component of the
binary (Blaauw 1961). Since we are interested in the production of
very massive ($\geq 60-70 \, \msun$) runaways, one should assume
that the primary star was a very massive star as well, with an
initial mass comparable to that of the secondary star. (Note that
the secondary star can increase its mass due to Roche-lobe
overflow of the primary star so that the mass of the runaway star
could be larger than the initial mass of the primary star.)

\begin{figure}
\includegraphics[width=8.5cm]{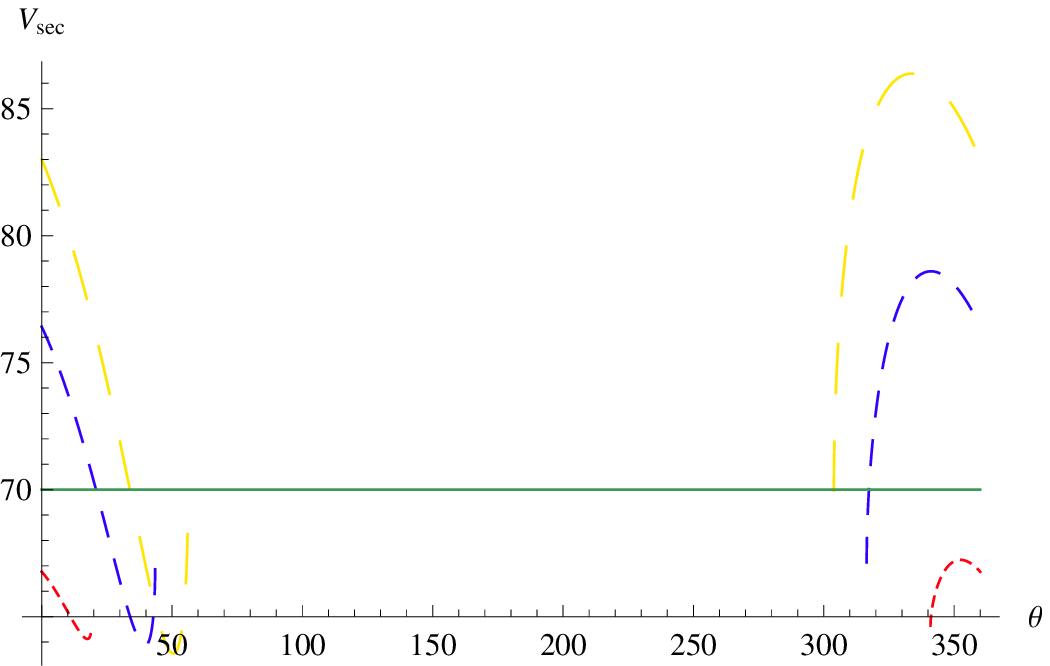}
\includegraphics[width=8.5cm]{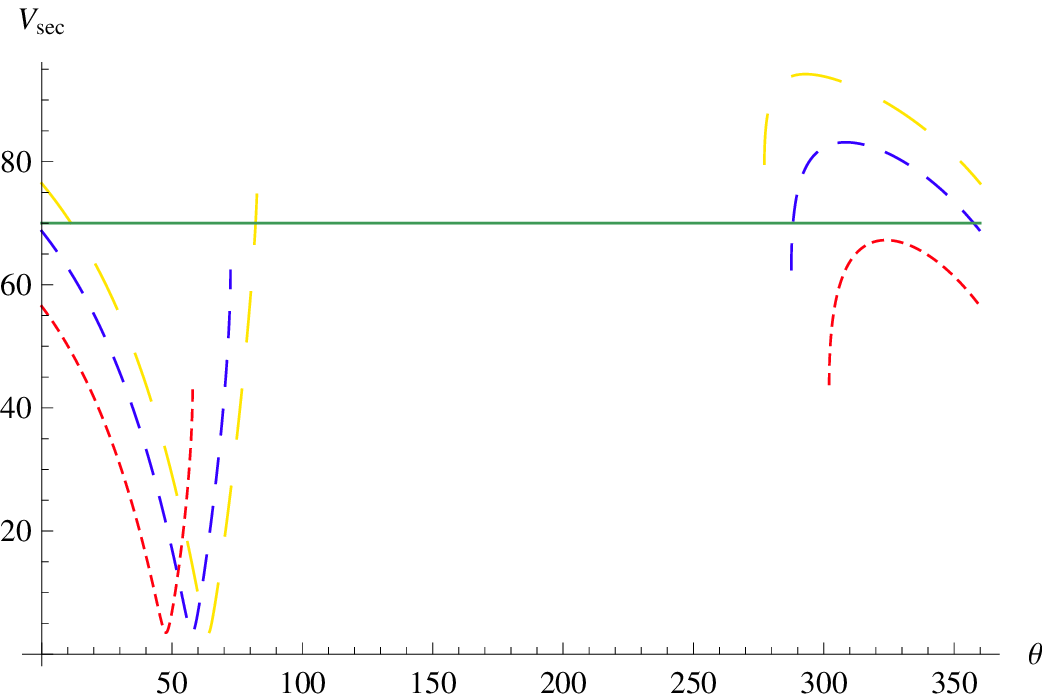}
\centering \caption{The peculiar velocity of a $70 \, \msun$
(secondary) star as a function of the angle, $\theta$, between the
kick vector and the direction of motion of the primary (exploding)
$15 \, \msun$ star and the magnitude of the kick, $w$, attained by
the stellar supernova remnant (black hole) of mass $M_{\rm BH}$.
{\it Upper panel}: $M_{\rm BH} =5 \, \msun$, $w=250 \, \kms$
(short-dashed line), $300 \, \kms$ (middle-dashed line), $350 \,
\kms$ (long-dashed line). {\it Bottom panel}: $M_{\rm BH} =10 \,
\msun$, $w=400 \, \kms$ (short-dashed line), $500 \, \kms$
(middle-dashed line), $600 \, \kms$ (long-dashed line). The
discontinuities in the curves correspond to a range of angles
$\theta$ for which the system remains bound. The horizontal line
indicates the peculiar velocity of the runaway star of $70 \,
\kms$. See text for details.} \label{fig:kick}
\end{figure}

Stellar evolutionary models suggest that the pre-supernova mass of
stars with initial (zero-age main-sequence) masses, $M_{\rm
ZAMS}$, from 12 to $120 \, \msun$ do not exceed $\sim 10-17 \,
\msun$ (Schaller et al. 1992; Vanbeveren, De Loore \& Van
Rensbergen 1998; Woosley, Heger \& Weaver 2002; Meynet \& Maeder
2003) and that it is maximum for stars with $M_{\rm ZAMS} \simeq
20-25 \, \msun$ and $\ga 80 \, \msun$ [see Fig.\,6 of Meynet \&
Maeder (2003)]. In the first case, the supernova explosion leaves
behind a neutron star, while in the second one the stellar
supernova remnant is a black hole of mass of $\simeq 5-10 \,
\msun$ (e.g. Woosley et al. 2002; Eldridge \& Tout 2004). It is
clear that a very massive binary cannot be unbound by a {\it
symmetric} supernova explosion since the system loses much less
than a half of its pre-supernova mass (Boersma 1961). One should
therefore assume that the stellar supernova remnant, a $5-10 \,
\msun$ black hole, received at birth a kick velocity exceeding the
escape velocity from the system (Stone 1982; Tauris \& Takens
1998). In this case, the peculiar velocity of the runaway star
strongly depends on the magnitude and the orientation of the kick
attained by the black hole (Tauris \& Takens 1998; Gvaramadze
2006a, 2009). Moreover, the runaway star can achieve the highest
velocity if the pre-supernova binary was as tight as possible,
i.e. if the secondary star of radius $r_2$ was close to filling
its Roche lobe, $r_2 \sim r_{\rm L}$, where $r_{\rm L}$ is the
radius of the Roche lobe, given by (Eggleton 1983)
\begin{equation}
r_{\rm L} = {0.49aq^{2/3} \over 0.6q^{2/3} +\ln (1+q^{1/3})} \, ,
\end{equation}
$a$ is the binary semimajor axis, $q=m_2 /m_1$, and $m_1$ and
$m_2$ are the masses of the primary and the secondary stars,
respectively. Assuming that the secondary is a $70 \, \msun$
main-sequence star of radius
\begin{equation}
r_2 = 0.8(m_2 /\msun)^{0.7}\rsun
\end{equation}
(Habets \& Heintze 1981) and adopting the maximum pre-supernova
mass of the exploding star of $15 \, \msun$, one finds from
equations (1) and (2) that $a\simeq 30 \, \rsun \, (\simeq
2r_{2})$. Then using equations (44)-(47) and (51)-(56) given in
Tauris \& Takens (1998), one finds that to produce a $70 \, \msun$
runaway star with a velocity of $\simeq 70 \, \kms$ the black hole
should attain a kick velocity of at least $\simeq 250 \, \kms$ (if
the mass of the black hole is $5 \, \msun$; see
Fig.\,\ref{fig:kick}) or $\simeq 400 \, \kms$ (if the mass of the
black hole is $10 \, \msun$), while the angle, $\theta$, between
the kick vector and the direction of motion of the exploding star
should be in a certain range (the smaller the kick the narrower
the range of allowed angles). Although one cannot exclude a
possibility that a $5-10 \, \msun$ black hole can attain a kick of
several hundreds of $\kms$ and of appropriate orientation (see
Gualandris et al. 2005 for an example of such case), we consider
the binary-supernova scenario as highly unlikely (cf. Gvaramadze
2007, 2009; Gvaramadze \& Bomans 2008a).

It is obvious that the binary-supernova scenario cannot be applied
to the runaway stars ejected from young ($\leq 2-3$ Myr) clusters.
The most massive stars in these clusters simply have no time to
end their lives in supernova explosions. Similarly, the young
($\sim 1-2$ Myr) ages of O2-type runaway stars in the LMC are also
inconsistent with the binary-supernova scenario (cf. Evans et al.
2010). Moreover, the large separations of the very massive runaway
stars (listed in Table\,\ref{tab:VMR}) from their plausible parent
clusters and OB associations imply that the majority of these
stars were ejected very soon after the birth, which also argues
against the binary-supernova scenario (Gvaramadze et al. 2010).

\section{Dynamical ejection scenario: massive runaway stars from three-body encounters}
\label{sec:hills}

An alternative to the binary-supernova scenario is the scenario based
on three- and four-body dynamical encounters in dense stellar systems
(Poveda et al. 1967; van Albada 1968; Aarseth 1974; Kroupa 1998;
Gualandris, Portegies Zwart \& Eggleton 2004; Pflamm-Altenburg \&
Kroupa 2006; Gvaramadze 2007, 2009; Gvaramadze \& Bomans 2008a;
Gvaramadze, Gualandris \& Portegies Zwart 2008, 2009). The best
studied and possibly the most efficient process responsible for the
origin of high-velocity stars is the close dynamical encounter between
two hard binary stars (Mikkola 1983; Leonard \& Duncan
1990). Numerical experiments performed by Leonard (1991) show that in
the course of binary-binary encounters one of the binary components
can occasionally be ejected with a velocity comparable to the escape
velocity from the surface of the most massive star in the binaries,
i.e. with a velocity of $\sim 1000 \, \kms$.

Three-body encounters can also produce very high-velocity runaway
stars, provided that the mass ratio of the single star to the mass
of the (hard) binary is either $>>$ or $<< 1$ (Hills \& Fullerton
1980). In the first case, one of the binary components could be
replaced by the very massive star in a so-called exchange
encounter, while the second component is ejected with a high
velocity (e.g. Gvaramadze et al. 2009). For equal mass binary
components and zero impact parameter, exchange encounters produce
a typical ejection velocity of $\sim 1.8 V_{\rm orb}$, where
$V_{\rm orb}$ is the orbital velocity of the ejected star in the
original binary (Hill \& Fullerton 1980). In the second case, the
low-mass star is scattered by the very massive binary and attains
a typical velocity of $\sim 0.8 V_{\rm orb}$ in a so-called fly-by
encounter (here $V_{\rm orb}$ is the orbital velocity in the very
massive binary with equal mass components). In both cases, the
ejected star gains its kinetic energy at the expense of the
increased binding energy of the post-encounter binary. In response
to the encounter, the binary recoils with a fraction $M_3/(M_1
+M_2 )$ of the velocity of the ejected star ($M_1 + M_2$ and $M_3$
are the masses of the post-encounter binary and the runaway star).

In the following, we will concentrate on dynamical encounters
between very massive binaries and single stars of mass comparable
to that of the binary components. For illustrative purposes, we
assume that the massive binaries have parameters similar to those
of the most massive known binary systems in the Galaxy, WR\,20a
and NGC\,3603-A1 (see Table\,\ref{tab:VMB}). Both systems
are very tight, with the semimajor axes of only $\simeq 3$ times
larger than the radii of the primary stars. The binding energy of
these binaries is comparable to the energy of a supernova
explosion, $\sim 10^{51}$ erg. If a $70 \, \msun$ intruding star
extracts only one per cent of this energy, it will attain a
peculiar velocity of $\sim 100 \, \kms$, which is large enough to
explain the observed peculiar velocities of the very massive
runaway stars (see Table\,\ref{tab:VMR}).
\begin{table}
  \caption{Most massive known Galactic binary stars.}
  \label{tab:VMB}
  \renewcommand{\footnoterule}{}
  \begin{center}
  \begin{minipage}{\textwidth}
\begin{tabular}{lcccccccccc}
\hline
Star & $M_1 (\msun) + M_2 (\msun$) & $a$ ($\rsun$) & References \\
\hline
WR\,20a & $83\pm 5+82\pm 5$ & $\simeq 55$ & 1,2 \\
NGC\,3603-A1 &  $116\pm 31+ 89\pm 16$ & $\simeq 60$ & 3 \\
\hline
\end{tabular}
 \end{minipage}
    \end{center}
(1) Bonanos et al. 2004; (2) Rauw et al. 2005; (3) Schnurr et al.
2008a.
%\end{minipage}
\end{table}

\section{Numerical experiments}
\label{sec:num}

In this section, we perform numerical simulations of three-body
encounters in order to obtain the velocity distribution for
runaway stars produced in the course of interactions between
(massive) single stars and a hard very massive binary. The
simulations are carried out with the {\tt sigma3} package included
in the STARLAB software environment (McMillan \& Hut 1996;
Portegies Zwart et al. 2001;
http://www.ids.ias.edu/{$\sim$starlab). The stars are treated as
point masses interacting gravitationally. However, the monitoring
of the relative distances between pairs of stars combined with the
information on the stellar radii allows us to identify collisions.
The stellar radii are determined via the mass-radius relationship
given by equation (2). If two stars come closer than the sum of
their radii the calculation is stopped and the encounter is
classified as a collision (merger).

We consider a target binary composed of stars of mass $M_1$ and
$M_2$, initial semimajor axis $a$ and eccentricity $e$, and an
intruding star of mass $M_3$ with an initial velocity $V_{\rm
rel}$ relative to the centre of mass of the binary. $V_{\rm rel}$
is set to $5\, \kms$, in accordance with typical dispersion
velocities in young massive clusters. The angles that define the
spatial orientation of the binary with respect to the single star
are randomized in a Monte Carlo fashion (see Hut \& Bahcall 1983).
The eccentricity of the binary is drawn from a thermal
distribution $P(e)=2e$ (Heggie 1975), having set a maximum value
in order to guarantee that the two binary components do not come
into contact at the first pericenter passage. The impact parameter
$b$ is randomized according to an equal probability distribution
for $b^2$ in the range $[0-b_{\rm max}]$. The maximum value
$b_{\rm max}$ is determined automatically for each experiment [see
Gualandris et al. (2004) for a description]. Energy conservation
is usually better than one part in $10^6$ and, in case the error
exceeds $10^{-5}$, the encounter is rejected. The accuracy in the
integrator is chosen in such a way that at most 5 per cent of the
encounters are rejected.

\subsection{WR\,20a-like binary}

In the first set of simulations we focus on interactions in which
a single star of mass $M_3$ (ranging from 3 to $80 \, \msun$)
encounters a WR\,20a-like binary ($M_1 = M_2 = 80 \, \msun$, $a=55
\, \rsun \simeq 0.25$ AU; cf. Table\,\ref{tab:VMB}).

In Fig.\,\ref{fig:branch} we present the probability of different
outcomes (branching ratios) as a function of $M_3$. For each value
of $M_3$ we perform a total of 3000 scattering experiments, which
result either in a fly-by or a merger. Ionizations never take
place as the binary is too hard to be dissociated by the incoming
star. The small binary separation (comparable to the radii of the
binary components) makes exchange encounters very rare so that
their contribution to the production of runaway stars is
negligible. The only encounters which can produce runaways are
fly-by encounters. The probability of these encounters decreases
with increasing $M_3$ and drops to $\simeq 20$ per cent for $M_3 =
80 \, \msun$. Correspondingly, the fraction of mergers increases
to $\simeq 80$ per cent.

\begin{figure}
\begin{center}
\includegraphics[width=6cm,angle=270,clip=]{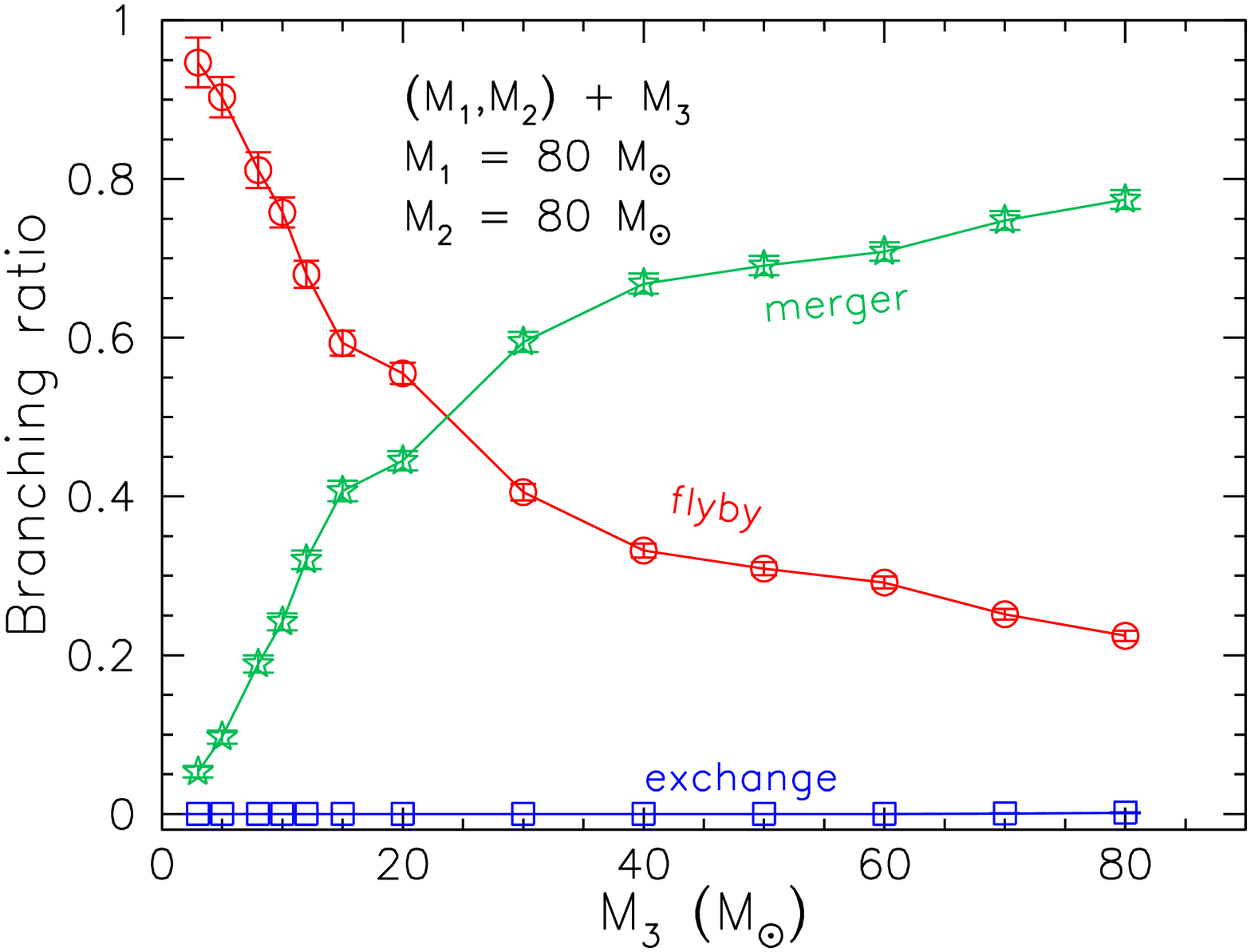}
\end{center}
\caption{Branching ratio for the outcome of encounters between a
  $(80,80)\, \msun$ binary and a single star as a function of its mass.
  The different outcomes are: merger (stars), fly-by (circles),
  and exchange (squares). The error bars represent
  the formal ($1\sigma$) Poissonian uncertainty of the measurement.}
\label{fig:branch}
\end{figure}
\begin{figure}
\begin{center}
\includegraphics[width=6cm,angle=270,clip=]{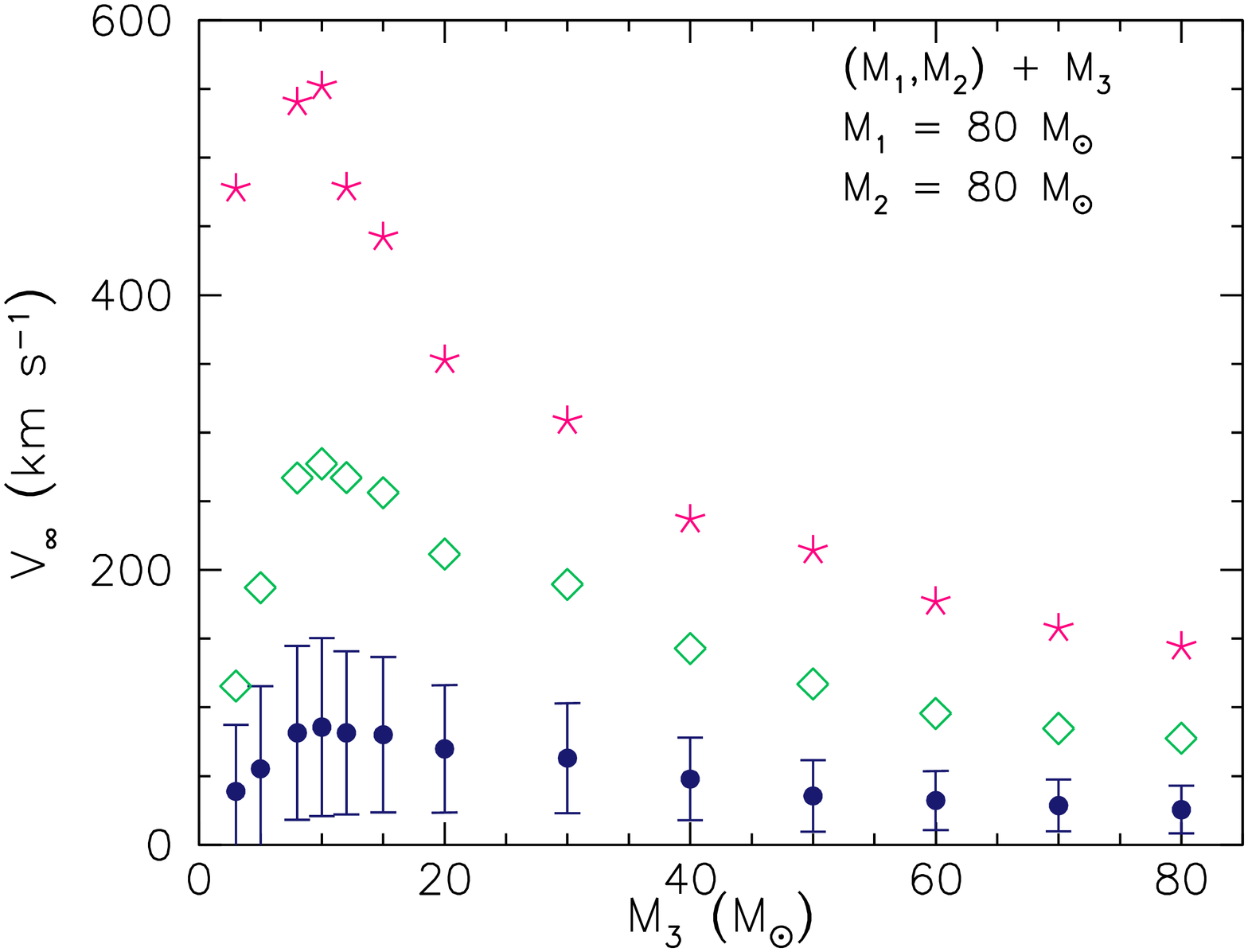}
\end{center}
\caption{Average velocity of escapers as a function of mass.  Circles
  represent the average velocity, diamonds indicate the velocity
  $V_{\rm max}$ for which 10 per cent of the encounters have
  $V_{\infty} > V_{\rm max}$, and stars indicate the velocity $V_{\rm
    max}$ for which 1 per cent of the encounters have $V_{\infty} >
  V_{\rm max}$. The error bars indicate the 1$\sigma$ deviation from
  the mean. For clarity, we only show them for one data set.}
\label{fig:velo}
\end{figure}
\begin{figure}
\begin{center}
\includegraphics[width=6cm,angle=270,clip=]{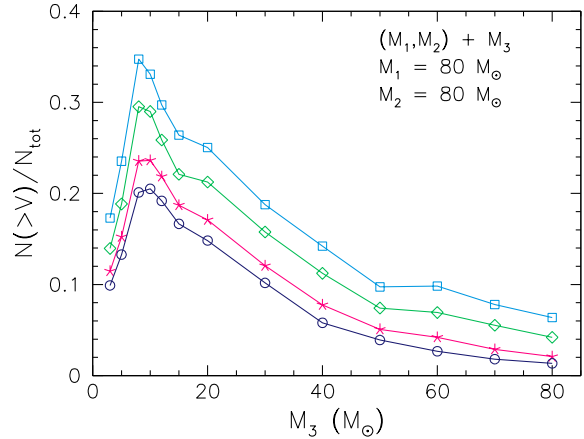}
\end{center}
\caption{The probability of fly-by encounters between a $(80,80)\,
\msun$ binary (with $a=0.25$ AU) and a single star (of mass $M_3$)
resulting in different ejection velocities: $> 30 \, \kms$
(squares), $> 50 \, \kms$ (diamonds), $> 80 \, \kms$ (stars), $>
100 \, \kms$ (circles).} \label{fig:brarun}
\end{figure}
\begin{figure}
\begin{center}
\includegraphics[width=8.5cm,angle=0,clip=]{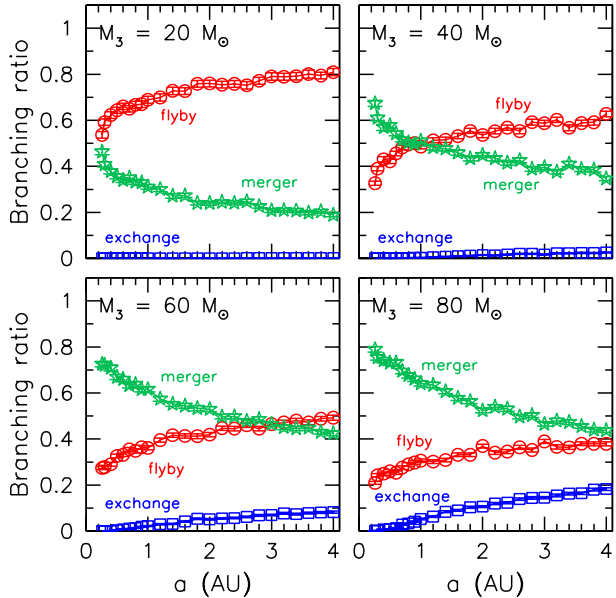}
\end{center}
\caption{Branching ratios for the outcomes of encounters between a
  $(80,80)\, \msun$ binary and a single star of mass $M_3 =20,40,60$
  and $80 \, \msun$ as a function of the binary semimajor axis.
  The different outcomes are: merger (stars),
  fly-by (circles), and exchange (squares). The error bars represent
  the formal ($1\sigma$) Poissonian uncertainty of the measurement.}
\label{fig:brasemi}
\end{figure}

Fig.\,\ref{fig:velo} shows the average velocity of escapers
produced in fly-by encounters as a function of their mass, $M_3$.
It can be seen that the velocity is a non-monotonic function of
$M_3$: it rapidly grows to its maximum value at $M_3 =10 \, \msun$
and gradually decreases afterwards. This behaviour is the result
of the interplay between two competing factors: (i) the more
massive the intruding star the greater its ability to affect the
binding energy of the binary system and thereby to increase the
kinetic energy of the system, and (ii) the more massive the
intruding star the larger the recoil velocity of the binary and
the smaller the velocity of the ejected star relative to the
centre of mass. Fig.\,\ref{fig:velo} also shows that the average
velocity attained by the $70-80 \, \, \msun$ stars is quite
moderate, $< 30 \, \kms$, so that they cannot be formally
classified as runaways. On the other hand, in 10 per cent of the
fly-by encounters the $70-80 \, \, \msun$ stars attain velocities
of $> 70 \, \kms$, and occasionally (in 1 per cent of the fly-by
encounters) can be accelerated to even larger ($> 150 \, \kms$)
velocities.

In Fig.\,\ref{fig:brarun} we show the probability of fly-by
encounters resulting in ejection velocities from 30 to $100 \,
\kms$ (top to bottom). For $M_3 \ga 70 \, \msun$ about $6-8$ per
cent of all encounters produce escapers with peculiar velocities
of $> 30 \, \kms$, i.e. typical of runaway stars, while the higher
velocities ($> 80 \, \kms$) can be attained in $\simeq 2-3$ per
cent of the encounters. The percentage of high-velocity ($> 100 \,
\kms$) runaway OB stars ($M_3 \geq 8 \, \msun$) increases with
decrease of $M_3$ and reaches a maximum ($\simeq 20$ per cent) for
$M_3 \simeq 10 \, \msun$.

To study the effect of the initial semimajor axis of the very
massive binary, we performed further scattering experiments.
Fig.\,\ref{fig:brasemi} shows the branching ratios as function of
$a$ for four different values of the mass of the intruding star,
$M_3 =20,40,60$ and $80 \, \msun$. One can see that the larger the
semimajor axis the larger the percentage of fly-by encounters.
Fig.\,\ref{fig:brasemi} also shows that the wider the binary and
the more massive the intruding star the larger the percentage of
exchange encounters, which for $M_3 =80 \, \msun$ and $a=4$ AU
reaches $\simeq 20$ per cent.

In Fig.\,\ref{fig:velosemi} we show the average velocity of
escapers as a function of $a$. One can see that the velocity
increases with increasing $a$ and then (for $a \ga 0.4-0.6$ AU)
gradually decreases. This counterintuitive growth can be
understood if one takes into account that the percentage of fly-by
encounters grows with $a$ as well (Fig.\,\ref{fig:brasemi}), so
that for a range of semimajor axes (up to several AU) the
increasing number of encounters producing runaway stars
compensates and even overcomes the velocity decrease caused by the
increase of the semimajor axis. Particularly, one can see that the
$80 \, \msun$ stars attain velocity of $\sim 100-140 \, \kms$ in
about 2-4 per cent of all encounters, if the binary separation is
$\simeq 0.3-3$ AU. Thus, three-body encounters involving binaries
with the mass of WR\,20a and the semimajor axes up to several AU
are quite efficient in producing very massive runaway stars.

\begin{figure}
\begin{center}
\includegraphics[width=6cm,angle=270,clip=]{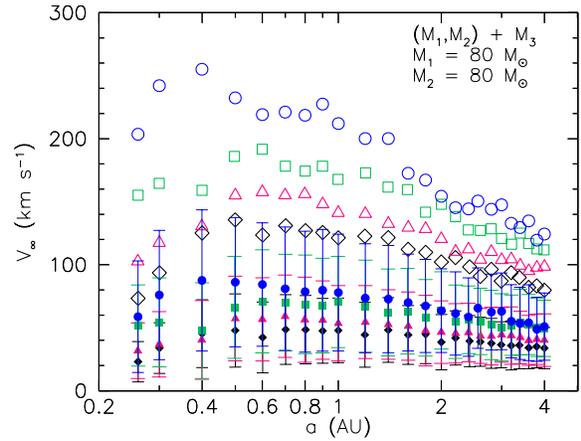}
\end{center}
\caption{Average velocity of escapers as a function of the initial
  binary semimajor axis in the interaction of a $(80,80) \, \msun$
  binary star with a single star of different mass: $M_3 =20 \, \msun$
  (circles), $M_3 =40 \, \msun$ (squares), $M_3 = 60 \, \msun$
  (triangles), $M_3 = 80 \, \msun$ (diamonds). Solid symbols represent
  the average velocity while the empty symbols indicate the velocity
  $V_{\rm max}$ for which 10 per cent of the encounters have
  $V_{\infty} > V_{\rm max}$. The error bars indicate the 1$\sigma$
  deviation from the mean. For clarity, we only show them for one data
  set.} \label{fig:velosemi}
\end{figure}

\subsection{NGC\,3603-A1-like binary}

\begin{figure}
\begin{center}
\includegraphics[width=6cm,angle=270,clip=]{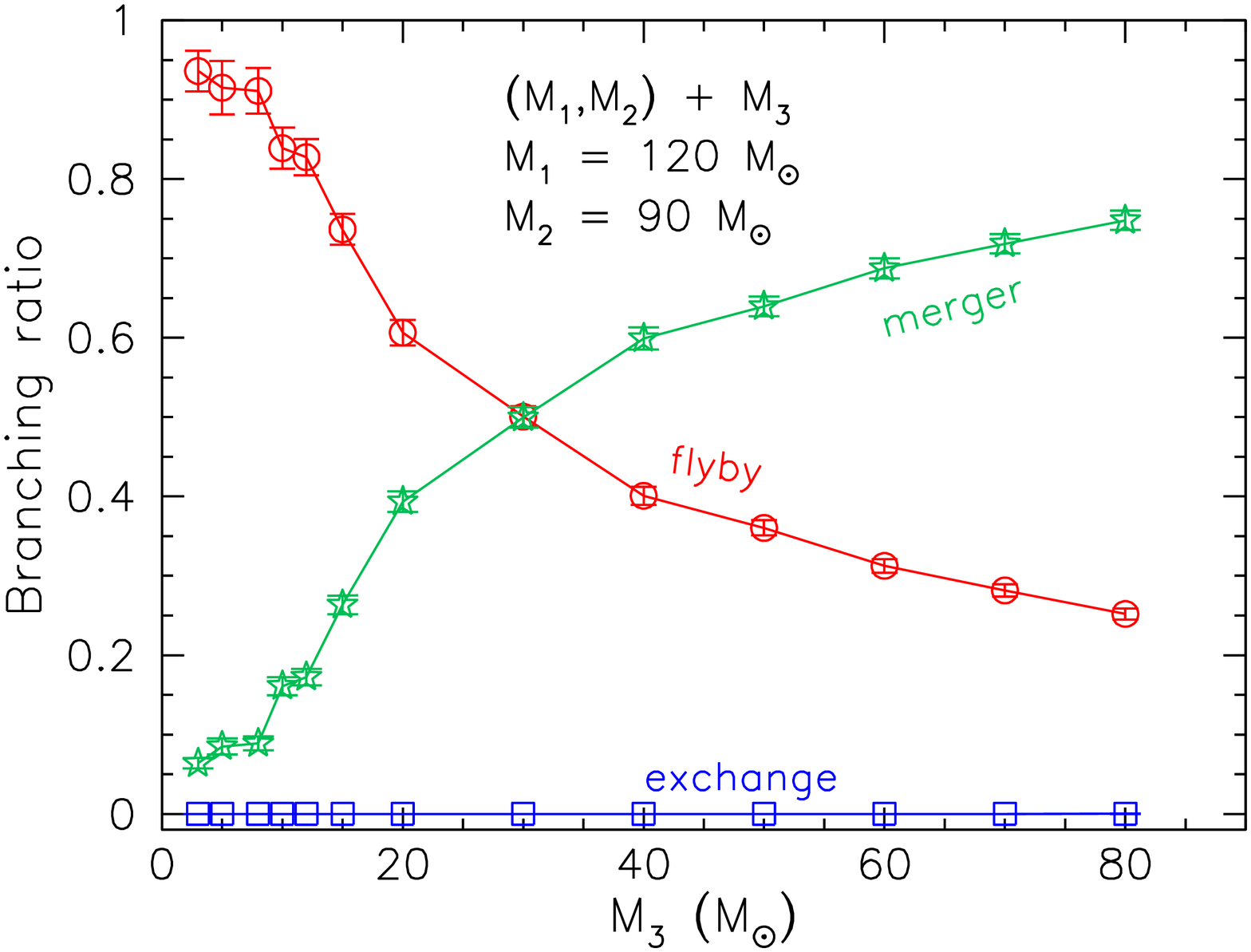}
\end{center}
\caption{Branching ratio for the outcome of encounters between a
  $(120,90)\, \msun$ binary and a single star as a function of its mass.
  The different outcomes are: merger (stars), fly-by (circles),
  and exchange (squares). The error bars represent
  the formal ($1\sigma$) Poissonian uncertainty of the measurement.}
\label{fig:branch-2}
\end{figure}
\begin{figure}
\begin{center}
\includegraphics[width=6cm,angle=270,clip=]{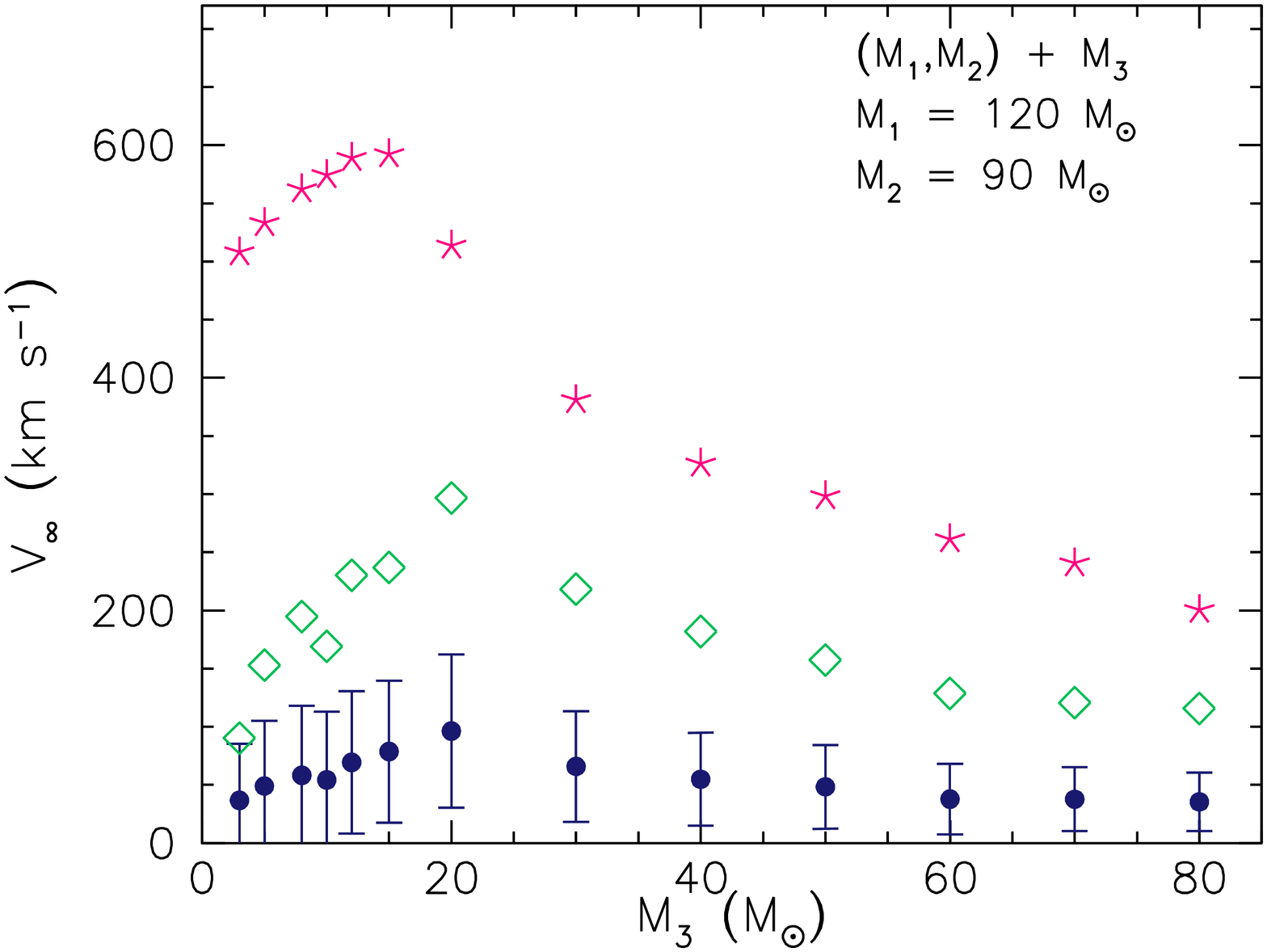}
\end{center}
\caption{Average velocity of escapers as a function of the mass.
  Circles represent the average velocity, diamonds indicate the
  velocity $V_{\rm max}$ for which 10 per cent of the encounters have
  $V_{\infty} > V_{\rm max}$, and stars indicate the velocity $V_{\rm
    max}$ for which 1 per cent of the encounters have $V_{\infty} >
  V_{\rm max}$. The error bars indicate the 1$\sigma$ deviation from
  the mean. For clarity, we only show them for one data set.}
\label{fig:velo-2}
\end{figure}
\begin{figure}
\begin{center}
\includegraphics[width=8.5cm,angle=0,clip=]{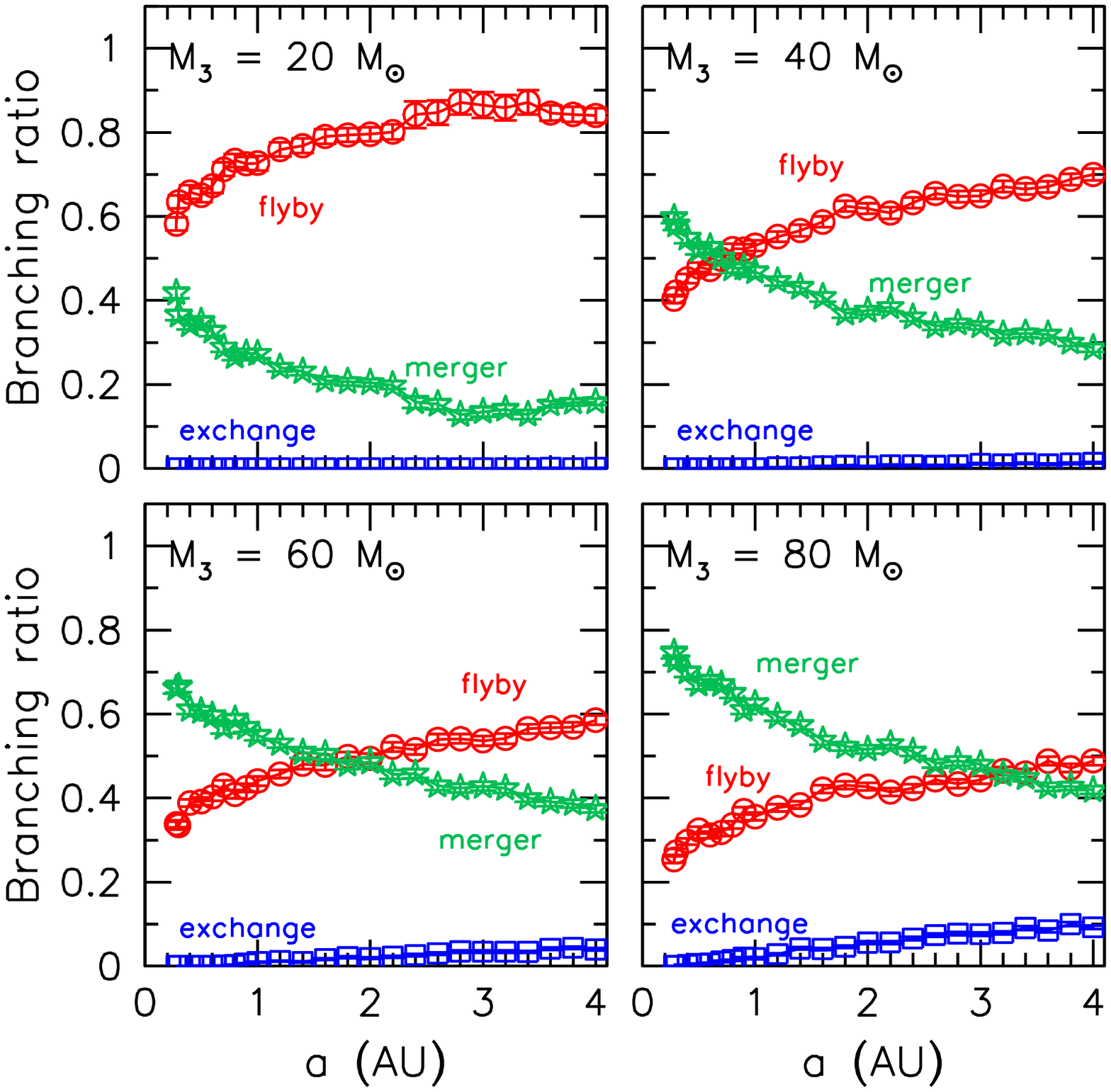}
\end{center}
\caption{Branching ratios for the outcomes of encounters between a
  $(120,90)\, \msun$ binary and a single star of mass $M_3 =20,40,60$
  and $80 \, \msun$ as a function of the binary semimajor axis.
  The different outcomes are: merger (stars), fly-by (circles), and
  exchange (squares). The error bars represent the formal ($1\sigma$)
  Poissonian uncertainty of the measurement.}
\label{fig:brasemi-2}
\end{figure}
\begin{figure}
\begin{center}
\includegraphics[width=6cm,angle=270,clip=]{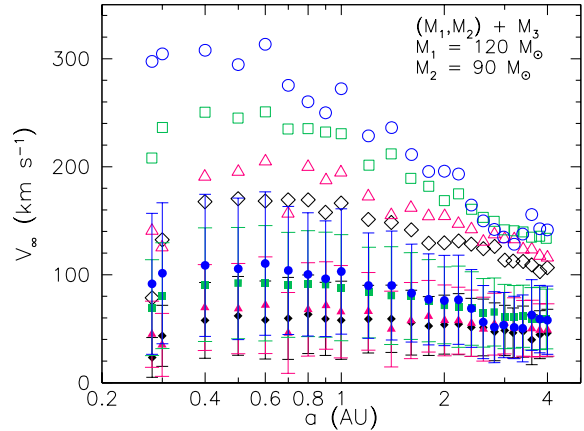}
\end{center}
\caption{Average velocity of escapers as a function of the initial
  binary semimajor axis in the interaction of a $(120,90) \, \msun$
  binary star with a single star of different mass: $M_3 =20 \, \msun$
  (circles), $M_3 =40 \, \msun$ (squares), $M_3 = 60 \, \msun$
  (triangles), $M_3 = 80 \, \msun$ (diamonds). Solid symbols represent
  the average velocity while the empty symbols indicate the velocity
  $V_{\rm max}$ for which 10 per cent of the encounters have
  $V_{\infty} > V_{\rm max}$. The error bars indicate the 1$\sigma$
  deviation from the mean. For clarity, we only show them for one data
  set.} \label{fig:velosemi-2}
\end{figure}

In the second set of simulations we consider three-body encounters
involving a NGC\,3603-A1-like binary ($M_1 =120 \, \msun , M_2 =90
\, \msun$, $a=60 \, \rsun \simeq 0.3$ AU; cf.
Table\,\ref{tab:VMB}). The branching ratios for these encounters
(Fig.\,\ref{fig:branch-2}) are almost identical to those given in
Fig.\,\ref{fig:branch}. The only difference is the somewhat larger
percentage of fly-by encounters, which is due to the larger
semimajor axis of the binary.

The average velocity of escapers is shown in
Fig.\,\ref{fig:velo-2}. As expected, the larger mass of the binary
results in a higher velocities of the escapers.
Fig.\,\ref{fig:velo-2} shows that in 10 per cent of the fly-by
encounters (or in $\simeq 2-3$ per cent of all encounters) the
$70-80 \, \msun$ stars attain velocities exceeding $110-120 \,
\kms$. One can see also that the NGC\,3603-A1-like binary is most
efficient at accelerating the $20 \, \msun$ stars, whose average
velocity is $\simeq 100 \, \kms$, while in about 6 per cent of all
encounters these stars attain velocities of $\geq 320 \, \kms$.

Fig.\,\ref{fig:brasemi-2} shows that the wider the very massive
binary the more frequent are the fly-by and the exchange
encounters. Correspondingly, the average velocity of escapers
grows with $a$ and after reaching the maximum value at $a\simeq
0.4-0.6$ AU it gradually decreases (see
Fig.\,\ref{fig:velosemi-2}). Still, for binaries as wide as
several AU, the average velocity of escapers remains higher than
that attained in the encounters with the NGC\,3603-A1-like binary.

\section{Discussion}
\label{sec:dis}

We performed numerical simulations of dynamical encounters between
very massive hard binaries and a single (massive) star in order to
explore the possibility that this three-body process is
responsible for the origin of very massive ($\geq 60-70 \, \msun$)
runaway stars, whose high peculiar velocities ($\geq 70 \, {\rm
km} \, {\rm s}^{-1}$) cannot be easily produced by the
disintegration of a binary system following a supernova explosion.
For illustrative purposes, we considered encounters with binaries
whose parameters are similar to those of the two most massive
known Galactic binaries, WR\,20a and NGC\,3603-A1 (see
Table\,\ref{tab:VMB}). Our simulations were motivated by the
observational fact that one of these binaries (WR\,20a) is
significantly offset ($\sim 1$ pc) from the centre of the parent
cluster (Westerlund\,2), which strongly suggests that the binary
was involved in a dynamical encounter with another massive star
and thereby was kicked out of the cluster (cf. Rauw et al. 2005).
We estimated the typical velocities produced in encounters between
very massive binaries and single stars and found that $\simeq 10$
per cent of the fly-by encounters (or $\simeq 2$ per cent of all
encounters) between the WR\,20a-like binary and a $70-80 \, \msun$
star produce escapers with velocities ($> 70 \, \kms$) similar to
those of the very massive Galactic runaway stars, $\lambda$\,Cep
and BD+43$\degr$\,3654. We also found that in about 2 per cent of
all encounters between the NGC\,3603-A1-like binary and a $80 \,
\msun$ star the escaper attains a velocity of $\ga 120 \, \kms$,
which is comparable to that of the most massive ($\sim 90 \,
\msun$) runaway star in the LMC, 30\,Dor\,016 (Evans et al. 2010;
see also Table\,1). The ejection velocities could be even higher
if the semimajor axes of the very massive binaries were larger
than those of WR\,20a and NGC\,3603-A1. In about 2-5 per cent of
encounters involving binaries with semimajor axes in the range
from 0.3 to $\sim 4$ AU, the ejection velocity of $80 \, \msun$
stars is $\geq 100-160 \, \kms$. We therefore argue that the
origin of (at least) some very massive high-velocity runaway stars
in the Galaxy and the LMC is associated with dynamical three-body
encounters.

Production of high-velocity massive stars via three-body
encounters is accompanied by recoil of the very massive binary in
the opposite direction to the ejected star. If the very massive
runaway star 30\,Dor\,016 were ejected in the field via the
three-body encounter in the central cluster, R136, of the
30\,Doradus nebula, then one would expect to find a very massive
binary on the opposite side of the cluster. Interestingly, such a
binary does indeed exist. The very massive binary R145
(HD\,269928), whose mass is of the same order of magnitude as
those of WR\,20a and NGC\,3603-A1 [Schnurr et al. (2009; also
private communication)], is located at $\simeq 1\farcm3$ (or
$\simeq 19$ pc in projection) from R136, just on the opposite side
of 30\,Dor\,016 (see Fig.\,\ref{fig:30Dor}\footnote{The image,
obtained in the framework of the {\it Spitzer} Survey of the Large
Magellanic Cloud (Meixner et al. 2006), was retrieved from the
NASA/IPAC Infrared Science Archive
(http://irsa.ipac.caltech.edu).}). If one assumes that
30\,Dor\,016 and R145 were ejected from R136 owing to the same
three-body encounter, then the conservation of the linear momentum
implies that the mass of the binary should be $\simeq 570 \,
\msun$, which is too large to be realistic (see Schnurr et al.
2009). From this it follows that either another very massive
binary exists at a larger distance from R136 or 30\,Dor\,016
attained its peculiar velocity in the course of a binary-binary
encounter (cf.  Gvaramadze \& Bomans 2008a). In the latter case,
30\,Dor\,016 and other stars involved in the encounter should not
lie on the same line.

Similarly, the large offset of R145 from R136 could be interpreted
as an indication that the binary was involved in an energetic
gravitational interaction in the parent cluster and that a massive
runaway star was ejected in the opposite direction.
Fig.\,\ref{fig:30Dor} shows that the B2 star (Rousseau et al.
1978) Sk\,$-$69$\degr$206 could be such a runaway. This star,
located $\simeq 17\arcmin$ to the west of R136, was identified as
a runaway via detection of its associated bow shock, whose
orientation is consistent with the possibility that
Sk\,-69$\degr$206 was ejected from 30\,Doradus (Gvaramadze et al.
2010). Assuming that the mass of Sk\,$-$69$\degr$206 is $\sim
10-15 \, \msun$\footnote{Note that Rousseau et al. (1978) give an
approximate spectral classification of Sk\,$-$69$\degr$206.}, one
finds that R145 should be as massive as $130-200 \, \msun$, which
is consistent with the mass estimate given in Schnurr et al.
(2009; also private communication).

The presence of numerous very massive (O2-3 and WN6h) stars spread
all around 30\,Doradus suggests that despite the young age (1-2
Myr) of R136, the cluster has already experienced a violent
dynamical evolution during which it lost a significant fraction of
its massive (single and binary) stars (cf. Brandl et al. 2007; see
also Pflamm-Altenburg \& Kroupa 2006; Moeckel \& Bate 2010). We
therefore predict that some of the very massive stars in the
30\,Doradus region are binary systems recoiled from R136 due to
three-body encounters in the cluster's core. The spectroscopic
monitoring of the brightest stars around 30\,Doradus would allow
to reveal the radial velocity variability and thereby to identify
binaries among them (e.g. Schnurr et al. 2008b), while the future
proper motion measurements for these stars with the space
astrometry mission {\it Gaia} will allow to determine the timing
of their ejection and thereby to link several stars to the same
ejection event.

\begin{figure}
\begin{center}
\includegraphics[width=8.3cm,angle=0,clip=]{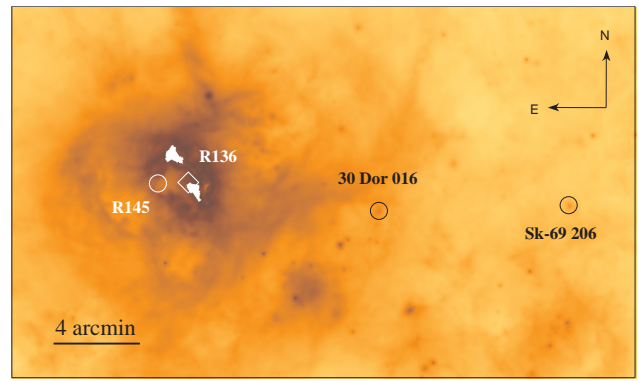}
\end{center}
\caption{{\it Spitzer} $24 \, \mu$m image of the 30\,Doradus
star-forming complex with position of its central cluster, R136,
marked by the diamond point. The positions of the very massive
binary R145 and two runaway stars, 30\,Dor\,016 and
Sk\,-69$\degr$206, are marked by circles. See text for details.}
\label{fig:30Dor}
\end{figure}

Our simulations also showed that the dynamical three-body
encounters provide an efficient channel for production of
high-velocity early B-type stars -- the progenitors of the
majority of neutron stars (pulsars). We found that $\simeq 6-8$
per cent of all encounters between $10-20 \, \msun$ single stars
and very massive binaries produce escapers with velocities ($>
200-350 \, \kms$) typical of pulsars (e.g. Hobbs et al. 2005), and
thereby could contribute to the origin of peculiar velocities of
these objects (cf. Gvaramadze 2006b, 2007; Gvaramadze et al. 2008;
Gvaramadze \& Bomans 2008b).

We also found that in $\simeq 1$ per cent of encounters involving
early B-type stars, the ejected star attains a velocity of $\geq
550-600 \, \kms$. It is worth noting that the velocity of this
order of magnitude was measured for the $11\pm 1 \, \msun$ star
HD\,271791 -- the fastest known massive runaway in the Galaxy
(Heber et al. 2008). It was shown by Gvaramadze (2009, 2010) that
the most likely origin of this extremely high-velocity star is
through dynamical interaction in the dense core of the parent star
cluster. Three-body encounters discussed in the present paper
could be one of the possible dynamical processes responsible for
the origin of HD\,271791.

A by-product of our simulations is the finding that about 1 per
cent of all encounters involving low-mass ($3-5 \, \msun$) stars
produces escapers with velocities of $> 500-600 \, \kms$, typical
of the so-called hypervelocity stars -- the ordinary stars moving
with velocities exceeding the Milky Way's escape velocity (Brown
et al. 2005). The existence of the hypervelocity stars was
foreseen by Hills (1988), who showed that a close encounter
between a tight binary and the supermassive black hole in the
Galactic Centre can produce escapers with a velocity of up to
several $1000 \, \kms$. An alternative explanation of the origin
of hypervelocity stars is that they attain extremely high
velocities via strong dynamical three- or four-body encounters in
the dense cores of massive star clusters located in the Galactic
disc (Gvaramadze 2006b, 2007, 2009; Gvaramadze et al. 2008, 2009)
or in the Large Magellanic Cloud (Gualandris \& Portegies Zwart
2007). The three-body encounters between single late B-type stars
and a very massive binary provide an additional channel for
production of the hypervelocity stars.

To conclude, we note that numerous uncertainties about the initial
conditions and early dynamical evolution of young star clusters
precludes us from making any estimates of the production rate of
high-velocity runaway stars (cf. Gvaramadze et al. 2008, 2009).

\section{Acknowledgements}
We are grateful to P.Kroupa, N.Langer, J.Pflamm-Altenburg,
O.Schnurr and T.M.Tauris for useful discussions, and to
P.P.Eggleton (the referee) for useful suggestions on the
manuscript. VVG acknowledges the Deutsche Forschungsgemeinschaft
for financial support. This research has made use of the NASA/IPAC
Infrared Science Archive, which is operated by the Jet Propulsion
Laboratory, California Institute of Technology, under contract
with the National Aeronautics and Space Administration.


\begin{thebibliography}{}
%
\bibitem{} Aarseth S.J., 1974, A\&A, 35, 237
\bibitem{} Aarseth S.J., Hills J.G., 1972, A\&A, 21, 255
\bibitem{} Blaauw A., 1961, Bull. Astron. Inst. Netherlands, 15, 265
\bibitem{} Blaauw A., 1993, in Massive Stars: Their Lives in the Interstellar Medium, ed. J.P.
Cassinelli, \& E.B. Churchwell, ASP Conf. Ser., 35, 207
\bibitem{} Boersma J., 1961, Bull. Astron. Inst. Netherlands, 15, 291
\bibitem{} Bonanos A.Z. et al., 2004, ApJ, 611, L33
\bibitem{} Brandl B.R., Portegies Zwart S.F., Moffat A.F.J., Chernoff D.F., 2007, in Massive Stars in Interactive Binaries,
ed. N. St.-Louis \& A.F.J. Moffat (San Francisco: ASP), 629
\bibitem{} Brown W.R., Geller M.J., Kenyon S.J., Kurtz M.J., 2005,
ApJ, 622, L33
\bibitem{} Clark J.S., Muno M.P., Negueruela I., Dougherty S.M., Crowther P.A., Goodwin S.P., de Grijs
R., 2008, A\&A, 477, 147
\bibitem{} Clarke C.J., Pringle J.E., 1992, MNRAS, 255, 423
\bibitem{} Comer\'{o}n F., Pasquali A., 2007, A\&A, 467, 23
\bibitem{} Conti P.S., Leep E.M., Lorre J.J., 1977, ApJ, 214, 759
\bibitem{} Danforth C.W., Chu, Y.-H., 2001, ApJ, 552, L155
\bibitem{} de Wit W.J., Testi L., Palla F., Zinnecker H., 2005, A\&A, 437, 247
\bibitem{} Eggleton P.P., 1983, ApJ, 268, 368
\bibitem{} Eldridge J.J., Tout C.A., 2004, MNRAS, 353, 87
\bibitem{} Evans C.J., Lennon D.J., Smartt S.J., Trundle C., 2006, 2006, A\&A, 456, 623
\bibitem{} Evans C.J., et al., 2010, ApJL, 715, L74
\bibitem{} Garc\'{i}a B., Mermilliod J.-C., 2001, A\&A, 368, 122
\bibitem{} Gies D.R., 1987, ApJS, 64, 545
\bibitem{} Gies D.R., Bolton C.T., 1986, ApJS, 61, 419
\bibitem{} Gualandris A., Portegies Zwart S., 2007, MNRAS, 376, L29
\bibitem{} Gualandris A., Portegies Zwart S., Eggleton P.P., 2004, MNRAS, 350, 615
\bibitem{} Gualandris A., Colpi, M., Portegie Zwart, S., Possenti, A.,  2005, ApJ, 618, 845
\bibitem{} Gvaramadze V.V., 2006a, A\&A, 454, 239
\bibitem{} Gvaramadze V.V., 2006b, in On the Present and Future of Pulsar Astronomy, 26th meeting
of the IAU, Joint Discussion 2, 16-17 August, 2006, Prague, Czech
Republic, JD02, \# 25
\bibitem{} Gvaramadze V.V., 2007, A\&A, 470, L9
\bibitem{} Gvaramadze V.V., 2009, MNRAS, 395, L85
\bibitem{} Gvaramadze V.V., 2010, in de Grijs R., Lepine J., eds, IAU Symp. 266, Star Clusters --
Basic Galactic Building Blocks throughout Time and Space,
Cambridge Univ. Press, Cambridge, p. 272
\bibitem{} Gvaramadze V.V., Bomans D.J., 2008a, A\&A, 485, L29
\bibitem{} Gvaramadze V.V., Bomans D.J., 2008b, A\&A, 490, 1071
\bibitem{} Gvaramadze V.V., Gualandris A., Portegies Zwart S., 2008, MNRAS, 385, 929
\bibitem{} Gvaramadze V.V., Gualandris A., Portegies Zwart S., 2009, MNRAS, 396, 570
\bibitem{} Gvaramadze V.V., Kroupa P., Pflamm-Altenburg J., 2010, A\&A, in press (astro-ph/1006.0225)
\bibitem{} Habets G.M.H.J., Heintze J.R.W., 1981, A\&AS, 46, 193
\bibitem{} Heber U., Edelmann H., Napiwotzki R., Altmann M., Scholz R.-D., 2008, A\&A, 483, L21
\bibitem{} Heggie D.C., 1975, MNRAS, 173, 729
\bibitem{} Hills J.G., 1975, AJ, 80, 809
\bibitem{} Hills J.G., 1988, Nat, 331, 687
\bibitem{} Hills J.G., Fullerton, L.W., 1980, AJ, 85, 1281
\bibitem{} Hobbs G., Lorimer D.R., Lyne A.G., Kramer M., 2005, MNRAS, 360, 974
\bibitem{} Hoogerwerf R., de Bruijne J.H.J., de Zeeuw P.T., 2001, A\&A, 365, 49
\bibitem{} Hut P., Bahcall J.N., 1983, ApJ, 268, 319
\bibitem{} Kiminki D.C., et al., 2007, ApJ, 664, 1102
\bibitem{} Kobulnicky H.A., Fryer C.L., 2007, ApJ, 670, 747
\bibitem{} Kobulnicky H.A., Gilbert I.J., Kiminki D.C., 2010, ApJ, 710, 549
\bibitem{} Kroupa P., 1998, MNRAS, 298, 231
\bibitem{} Lada C.J., Lada E.A., 2003, ARA\&A, 41, 57
\bibitem{} Larson R.B., 2001, in Zinnecker H., Mathieu R.D., eds, IAU Symp. 200, The Formation of Binary Stars,
Cambridge Univ. Press, Cambridge, p. 93
\bibitem{} Leonard P.J.T., 1991, AJ, 101, 562
\bibitem{} Leonard P.J.T., Duncan M.J., 1990, AJ, 99, 608
\bibitem{} Martins F., Schaerer D., Hillier D. J., 2005, A\&A, 436, 1049
\bibitem{} Mason B.D., Gies D.R., Hartkopf W.I., Bagnuolo W.G., ten Brummelaar T., McAlister
H.A., 1998, AJ, 115, 821
\bibitem{} Massey, P., Puls, J., Pauldrach, A.W.A., Bresolin, F., Kudritzki, R.P., Simon,
T., 2005, ApJ, 627, 477
\bibitem{} McMillan P.J., Binney J.J., 2010, MNRAS, 402, 934
\bibitem{} McMillan S.L.W., Hut P., 1996, ApJ, 467, 348
\bibitem{} Meixner M., et al., 2006, AJ, 132, 2268
\bibitem{} Mel'nik A.M., Dambis A.K., 2009, MNRAS, 400, 518
\bibitem{} Mermilliod J.-C., Garc\'{i}a B., 2001, in Zinnecker H., Mathieu R.D., eds, IAU Symp. 200, The Formation of Binary Stars,
Cambridge Univ. Press, Cambridge, p. 191
\bibitem{} Mikkola S., 1983, MNRAS, 203, 1107
\bibitem{} Meynet G., Maeder A., 2003, A\&A, 404, 975
\bibitem{} Moeckel N., Bate M.R., 2010, MNRAS, 404, 721
\bibitem{} Nota A. et al., 1994, in Clegg R.E.S., Stevens I.R., Miekle W.P.S., eds, Circumstellar Media in the Late Stages of Stellar Evolution, Cambridge Univ. Press,
Cambridge, p. 89
\bibitem{} Pflamm-Altenburg J., Kroupa P., 2006, MNRAS, 373, 295
\bibitem{} Pinsonneault M.H., Stanek K.Z., 2006, ApJ, 639, L67
\bibitem{} Portegies Zwart S.F., McMillan S.L.W., Hut P., Makino J. 2001, MNRAS, 321, 199
\bibitem{} Poveda A., Ruiz J., Allen C., 1967, Bol. Obs. Tonantzintla Tacubaya, 4, 86
\bibitem{} Preibisch T., Weigelt G., Zinnecker H., 2001, in Zinnecker H., Mathieu R.D., eds, IAU Symp. 200, The Formation of Binary Stars,
Cambridge Univ. Press, Cambridge, p. 69
\bibitem{} Rauw G. et al., 2005, A\&A, 432, 985
\bibitem{} Reid M.J. et al. 2009, ApJ, 700, 137
\bibitem{} Repolust T., Puls J., Herrero A., 2004, A\&A, 415, 349
\bibitem{} Rousseau J. et al. 1978, A\&AS, 31, 243
\bibitem{} Schaller G., Schaerer D., Meynet G., Maeder A., 1992, A\&AS, 96, 269
\bibitem{} Schilbach E., R\"{o}ser S., 2008, A\&A, 489, 105
\bibitem{} Schnurr O., Moffat A.F.J, St-Louis N., Casoli J., Chen\'{e} A.-N., 2008a, MNRAS, 389, L38
\bibitem{} Schnurr O., Moffat A.F.J, St-Louis N., Morrell N.I., Guerrero M.A., 2008b, MNRAS, 389, 806
\bibitem{} Schnurr O., Moffat A.F.J., Villar-Sbaffi A., St-Louis N., Morrell N.I., 2009, MNRAS, 395, 823
\bibitem{} Stone R.C., 1991, AJ, 102, 333
\bibitem{} Tauris T.M., Takens R.J., 1998, A\&A, 330, 1047
\bibitem{} van Albada T.S., 1968, Bull. Astron. Inst. Netherlands, 20, 57
\bibitem{} Van Buren D., McCray R., 1988, ApJ, 329, L93
\bibitem{} van Leeuwen F., 2007, A\&A, 474, 653
\bibitem{} Vanbeveren D., De Loore C., Van Rensbergen W., 1998, A\&AR, 9, 63
\bibitem{} Walborn N.R., 1973, AJ, 78, 1067
\bibitem{} Woosley S.E., Heger A., Weaver T.A., 2002, RvMP, 74, 1015
\bibitem{} Zinnecker H., Yorke H.W., 2007, ARA\&A, 45, 481

\end{thebibliography}
\end{document}